\documentclass{eas}
\usepackage{graphicx}
\newcommand{\hmpc}{h^{-1}{\rm Mpc}}
\newcommand{\kms}{\;{\rm km}\,{\rm s}^{-1}}
\newcommand{\CIV}{\hbox{C\,{\sc iv}}}
\newcommand{\CV}{\hbox{C\,{\sc v}}}
\newcommand{\CVI}{\hbox{C\,{\sc vi}}}
\newcommand{\OVI}{\hbox{O\,{\sc vi}}}
\newcommand{\gad}{{\sc Gadget-2}}
\newcommand{\vw}{{v_{\rm wind}}}

\begin{document}

%%-----------------------------
%%      the top matter
%%-----------------------------
\title{When Does the Intergalactic Medium Become Enriched?} 
\runningtitle{When Does the IGM Become Enriched?}
\author{Benjamin D. Oppenheimer}\address{Steward Observatory, 933 N. Cherry Ave., Tucson, AZ, USA}
\author{Romeel Dav\'e$^1$}
\author{Kristian Finlator$^1$}
\begin{abstract}

We use cosmological hydrodynamic simulations including galactic
feedback based on observations of local starbursts to find a
self-consistent evolutionary model capable of fitting the observations
of the intergalactic metallicity history as traced by \CIV~between
$z=6.0\rightarrow 1.5$.  Our main finding is that despite the relative
invariance in the measurement of $\Omega(\CIV)$ as well as the column
density and linewidth distributions over this range, continual
feedback from star formation-driven winds are able to reproduce the
observations, while an early enrichment scenario where a majority of
the metals are injected into the IGM at $z>6$ is disfavored.  The
constancy of the \CIV~observations results from a rising IGM
metallicity content balanced by a declining \CIV~ionization fraction
due to a 1) decreasing physical densities, 2) increasing ionization
background strength, and 3) metals becoming more shock-heated at lower
redshift.  Our models predict that $\sim20\times$ more metals are
injected into the IGM between $z=6\rightarrow2$ than at $z>6$.  We
show that the median \CIV~absorber at $z=2$ traces metals injected 1
Gyr earlier indicating that the typical metals traced by \CIV~are
neither from very early times nor from very recent feedback.

\end{abstract}
\maketitle
%%-----------------------------
%%      your text
%%-----------------------------
\section{Introduction}

Metals observed in the intergalactic medium (IGM) via quasar
absorption line spectroscopy are thought to originate from some form
of star formation-driven outflows as dynamical disruption is too
inefficient (Aguirre et al. \cite{agu01}).  When and how metals are
injected into the IGM are still a matter of debate.  Observations of
\CIV~in the IGM show surprisingly little evolution between $z\sim
5\rightarrow 2$ (Songaila \cite{son01}, Schaye et al. \cite{sch03}, \&
Songaila \cite{son05}), despite this being the age of peak star
formation.  These observations motivate early enrichment scenarios by
primeval galaxies and/or Populations-III stars at $z>6$ where physical
distances are smaller and the gravitational potential wells are
shallower.

In Oppenheimer \& Dav\'e (\cite{opp06}) (hereafter OD06), we explore a
variety of galactic outflow models in \gad~cosmological hydrodynamic
simulations with the specific emphasis on modeling the \CIV~absorption
observations between $z=6.0\rightarrow1.5$.  We are able to reproduce
the main \CIV~observations including the relative invariance of
$\Omega$(\CIV) by implementing galactic-scale outflows based on
observed local starbursts (Martin \cite{mar05}, Rupke et
al. \cite{rup05}).  Only a narrow range of wind parameters using the
physically-motivated momentum-driven wind scenario (Murray, Quartet,
\& Thompson \cite{mur05}) are able to enrich the IGM without
overheating it at early times.  Our models favor metal enrichment tied
to concurrent star formation, as opposed to an early enrichment
scenario with the majority of metals injected into the IGM before
$z=6$. Observations supporting this include the ubiquitous outflows
seen from star-forming galaxies at high-redshift galaxies (Pettini et
al. \cite{pet01}) and enhanced \CIV~and \OVI~observed around
Lyman-break galaxies at $z\sim 2-3$ (Adelberger et al. \cite{ade03},
\cite{ade05}).

In these proceedings, we provide further arguments that the majority
of metals in the IGM cannot be injected at very early times.  Instead,
we show the increasing IGM metallicity is balanced by an increasing
\CIV~ionization correction resulting in observations of $\Omega$(\CIV)
that remain relatively invariant from $z=6\rightarrow2$.  We further
explore the age of the metals in the IGM traced by \CIV~absorbers.

\section{Simulations}

We use the N-body+hydrodynamic code \gad\ (Springel \cite{spr05}) with
improvements including metal-line cooling and tunable superwind
feedback as described in \S2 of OD06.  Briefly, the simulations we
discuss here use an $\Omega=0.3$ $\Lambda$CDM cosmology in a 16
$\hmpc$ box represented with $2\times256^3$ particles, giving a gas
particle resolution of $3.9\times10^6 M_{\odot}$.  \S6 of OD06 shows
that this sufficiently resolves most \CIV~absorbers.

For simplicity, we focus on only two of the six outflow models of
OD06, the ``constant wind'' (cw) and the ``momentum-driven wind''
(vzw) models.  The cw model assumes $\vw=484$ $\kms$ and a mass
loading factor $\eta=2$, while the vzw model uses the relations
$\vw\propto\sigma$ and $\eta\propto\sigma^{-1}$ ($\sigma$ being the
galaxy velocity dispersion).  

Spectra were generated from the simulations using our software {\tt
specexbin}, which first calculates the physical properties along a
line of sight and then applies ionization fractions from CLOUDY tables
of an optically thin slab of gas with an impinging ionizing background
field from Haardt \& Madau (\cite{haa01}) including quasars and
star-forming galaxies.  The resulting spectra were analyzed using the
AutoVP package to generate Voigt-profile fits from which line
statistics could be derived and the summed $\Omega$(\CIV) was
calculated.

\section{Continual vs. Early Enrichment}

\begin{figure} 
\includegraphics[angle=-90,scale=.45]{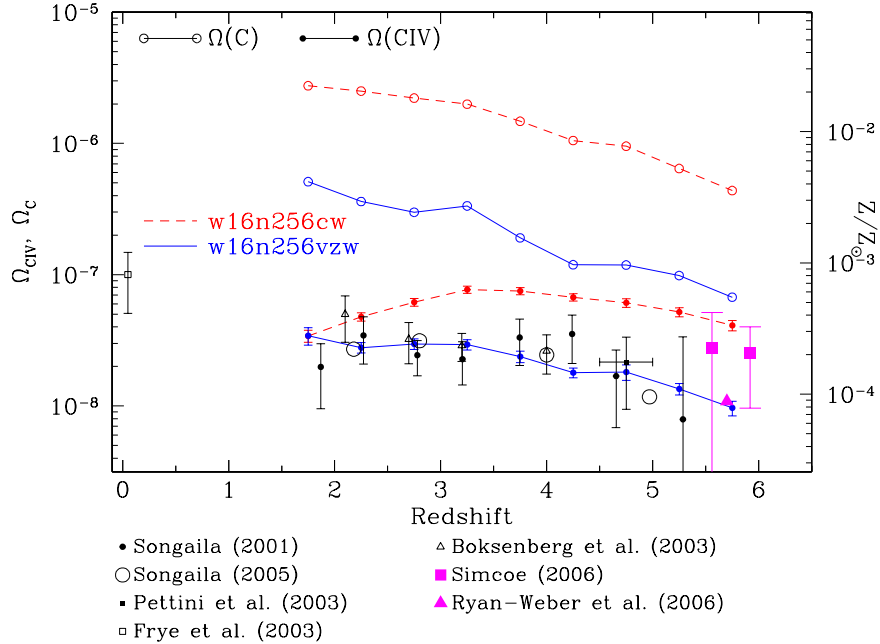}
\caption[]{Evolution of $\Omega$(\CIV) and $\Omega$(C), the mass
density of \CIV~and metals, respectively.  The constant wind (cw)
model is overly efficient at enriching the IGM early, and creates an
excessively hot IGM resulting in a much higher \CIV~ionization
correction that increases too sharply at later times.  The
momentum-driven wind (vzw) injects fewer metals in the IGM, but does
not require as extreme ionization correction due to less heating of
the IGM from slower wind velocities at earlier times.  The Ryan-Weber
et al. data point is a lower limit.  }
\label{fig1} 
\end{figure}

Figure \ref{fig1} displays the evolution of $\Omega$(\CIV) for a
variety of observations and our cw and vzw models.  A comparison to
$\Omega$(C) indicates significant ionization corrections, usually in
excess of 10 are necessary to derive the true metallicity from
$\Omega$(\CIV).  The ionization correction increases toward lower
redshift for three main reasons.  First and most importantly for the
vzw model, the physical densities decrease at lower redshift meaning
that \CIV~traces metals at higher overdensities, which occupy less
volume at lower redshift.  Second, the ionization background
strengthens between $z=6\rightarrow 2$ pushing the metals traced by
\CIV~to slightly higher overdensity.  Third, the metals become more
shock-heated at later times due to metal-enriched winds colliding with
increasingly faster infalling material toward low redshift.  Metals
are pushed into regions lower overdensity and higher temperature where
carbon becomes ionized into \CV~and \CVI.  This third effect is
greatest for the cw model causing its downturn in $\Omega$(\CIV) at
$z<3$.

New observations at $z>5.5$ (Ryan-Weber et al. \cite{rya06}, Simcoe
\cite{sim06}) may suggest even more \CIV~at high redshift than our
models produce.  The momentum-driven galactic superwind feedback
prescription based on observations of local starbursts and star
formation rates may not be enough to enrich the high-$z$ IGM.  We do
not rule out the possibility of another form of metal enrichment from
the earliest galaxies and stars, although we cannot conclude this
definitely.  Nevertheless, our point is to show that metal enrichment
from continual star formation is required to produce most of the
observed \CIV~by $z=2$.  

\begin{figure}
\includegraphics[scale=.25]{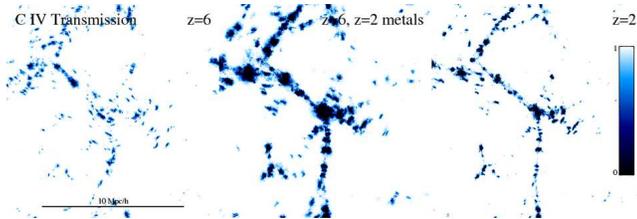}
\caption[]{These three panels show the \CIV~absorption in a 100 $\kms$
slice of the vzw model (15 $\hmpc$ across).  The left and right panels
show the metals at $z=6$ and $z=2$ respectively.  The middle panel
represents the early enrichment scenario where the metals in our
simulation at $z=2$ are placed at $z=6$, thus assuming metals
passively evolve from $z=6\rightarrow2$.  The \CIV~ionization
correction is much lower at $z=6$ resulting in $\sim4\times$ more
absorption at $z=6$ and an $\Omega$(\CIV) that must decline by a
factor of at least two from $z=6\rightarrow 2$, which is not observed.
We can therefore rule out that the majority of metals were injected
into the IGM before $z=6$.}
\label{fig2}
\end{figure}

As an exercise to show that the majority of metals cannot be injected
into the IGM at early times, we take the metals in our simulation at
$z=2$ and put them at $z=6$ densities and ionization conditions.  This
assumes that the metals passively evolve since $z=6$ without any new
metals being injected.  We also allow the IGM to cool to the $z=6$ IGM
equation of state (EoS) at $\delta<100$, thus assuming that the metals
were injected early enough to allow the IGM to cool.  Slices of
absorption in Figure \ref{fig2} show how this case (middle) would
result in much more absorption than the original $z=6$ or $z=2$ vzw
case (left \& right respectively).  The lower ionization corrections
at $z=6$ result in 4$\times$ more absorption leading to a predicted
drop in $\Omega$(\CIV) of at least 2$\times$.  Even if we leave the
metals at the $z=2$ shocked temperatures, there is still a significant
decline.  The fact that a constant $\Omega$(\CIV) is observed is
strong evidence that metals in the IGM must be continually injected
during the age of peak star formation.

\begin{figure}
\includegraphics[angle=-90,scale=.45]{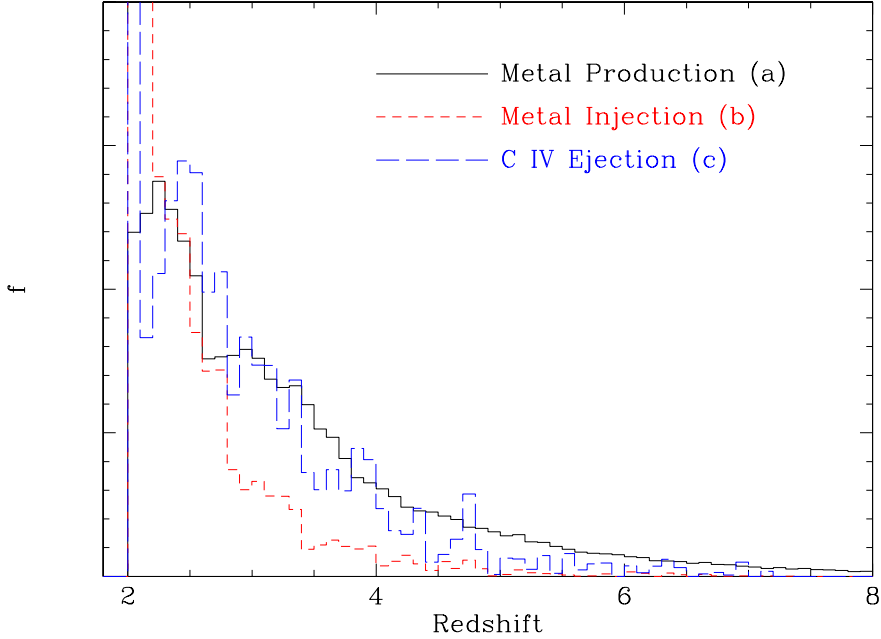}
\caption[]{The black histogram (a) represents the fraction of metals
produced at that redshift (proportional to the global star formation
rate) indicating that 95\% of metals at $z=2$ are generated at $z<6$
in the vzw model.  On the same scale, the red short-dashed histogram
(b) is the redshift when the metals were last injected into the IGM
via a wind, while the blue long-dashed histogram (c) is the same
except for metals traced by \CIV.  $\sim15$\% of \CIV~absorption at
$z=2$ arises from galactic/halo gas, which accounts for $\sim40$\% of
the gaseous metals.  The median \CIV~absorber traces metals injected 1
Gyr ago at $z=2$.  Thus \CIV~traces neither metals from only recent
star formation nor early enrichment, but instead from a variety of
ages.  All histograms are on the same scale to allow one to compare
them.}
\label{fig3}
\end{figure}

Finally, we explore the age of the typical \CIV~absorber at $z=2$.  We
track the histories of individual SPH particles in our hydrodynamical
simulation to see the redshift of their last injection into the IGM.
Figure 3 plots the redshift histogram of a) the metal production
(proportional to the global star formation rate), b) the last
injection into the IGM of those metals, and c) the same as (b) but for
metals traced by \CIV.  Although $\sim$40\% of the gaseous metals are
in galactic/halo gas, only 15\% of \CIV~absorption in spectra comes
from this gas leaving the other 85\% arising from the IGM.  The median
age of a \CIV~absorber (ie. the last time it was in a wind) is 1 Gyr
at $z=2$.  \CIV~traces metals injected at a variety of ages, and does
not exclusively trace metals from very recent star formation nor
metals from early enrichment.  The average particle is injected into
the IGM via a wind more than twice by $z=2$, indicating there is
significant recycling of metals in and out of the IGM.  Older metals
lie in lower overdensities suggesting observations that indicate
\CIV~resides at a variety of overdensities (ie. the pixel optical
depth method) require that metals be injected into the IGM at a
variety of redshifts.

\section{Conclusions}

Our cosmological hydrodynamic simulations using momentum-driven winds
from star-forming galaxies fit the main \CIV~observables between
$z=6.0\rightarrow1.5$ as described in OD06.  We further emphasize: 
\\
\\
$\bullet$ The $\Omega$(\CIV) measurement remains relatively invariant
after $z=6$ despite a vast majority ($>$90\%) of the IGM becoming
enriched between $z=6\rightarrow2$.  The ionization correction of
\CIV~decreases due to 1) decreasing physical densities, 2) increasing
ionization background strength, and 3) metals becoming more
shock-heated at lower redshift.
\\
\\
$\bullet$ We disfavor an early enrichment-only scenario,
because it would lead to a decreasing $\Omega$(\CIV) with time due to
the increasing \CIV~ionization correction.  The IGM is primarily
enriched via continual outflows from star-forming galaxies, although a
different form of metal enrichment from the first stars/galaxies may
be required for the $z>5.5$ \CIV~observations.
\\
\\
$\bullet$ \CIV~traces metals injected into the IGM at a variety of
ages, with the median age of 1 Gyr at $z=2$.  Only 15\% of \CIV~arises
from galactic/halo gas, while the other 85\% is from the IGM.  The
spread of ages of IGM metals may be necessary to enrich the variety of
overdensities as indicated by observations.

%%-----------------------------
%%      your bibliography
%%-----------------------------

\end{document}